\documentclass[12pt]{article}
\usepackage{amsmath}

\begin{document}
\begin{titlepage}

\renewcommand{\thefootnote}{\fnsymbol{footnote}}

\hfill TUW--99--02

\hfill Vers. 2.1 \\

\begin{center}
\vspace{0.5cm}

{\Large\bf Absolute conservation law for black holes}
\vspace{1.0cm}

{\bf D.\ Grumiller\footnotemark[1] and  
W. Kummer\footnotemark[2]
}
\vspace{7ex}

{Institut f\"ur
    Theoretische Physik \\ Technische Universit\"at Wien \\ Wiedner
    Hauptstr.  8--10, A-1040 Wien \\ Austria}
\vspace{1.5cm}

\footnotetext[1]{e-mail: {\tt grumil@hep.itp.tuwien.ac.at}}
\footnotetext[2]{e-mail: {\tt wkummer@tph.tuwien.ac.at}}

\end{center}

\begin{abstract}
In all 2d theories of gravity 
a conservation law connects the (space-time dependent) mass aspect function 
at all times and all radii with 
an integral of the matter fields. It depends on an arbitrary 
constant which may be interpreted as determining the initial 
value together with the initial values for the matter field. 
We discuss this for spherically reduced Einstein-gravity in a diagonal metric 
and in a Bondi-Sachs metric using the first order formulation of 
spherically reduced gravity, which allows easy and direct fixations of 
any type of gauge. The relation of our conserved quantity to the ADM and
Bondi mass is investigated. Further possible applications 
(ideal fluid, black holes in higher dimensions or AdS spacetimes etc.) are 
straightforward generalizations. 
\end{abstract}


\vfill
\end{titlepage}

\section{Introduction}
The treatment of two-dimensional models of gravity is simpler and 
more transparent in terms of a light-cone gauge for the Cartan 
variables, or, equivalently, in Eddington-Finkelstein (EF) gauge 
for the 2d metric \cite{KS92}, especially when these approaches are
combined with the first order formulation 
\cite{IK93,Str94,Scha94,KW95,KS96}.

Of course using traditional approaches -- e.g. performing the calculations in conformal gauge  \cite{LM94,MAN91} -- one must be able to obtain the same results. In practice, however, the resulting equations of motion are rather 
complicated. Therefore, it is not surprising that also a complete 
discussion of the global solution of such models has turned out 
to be easier in the EF gauge \cite{KS96/2} than in the 
corresponding previous studies of the conformal gauge 
\cite{Kat93} for the model of Katanaev and Volovich \cite{Kat86}. 
Based upon the use of the EF metric in a first order 
formulation also substantial progress has been 
achieved in the meantime regarding the quantum theory of such models 
\cite{KS92/2}.

The purpose of our present note is to draw attention to one 
particular aspect of the first order approach to spherically 
reduced gravity, namely the existence of an absolute (space and time)  
conservation law connecting the effective (space-time) dependent 
mass aspect function with the interacting matter fields \cite{KW95,KT98}. This 
conservation law generalizes some previously known special
cases, like e.g. the case without matter \cite{bankslough} and the one with
nondynamical matter \cite{mann}.
The differential equations describing 
spherically symmetric Einstein gravity interacting with matter 
are known for a long time. Many recent extensive numerical 
simulations of black hole formation \cite{Chop86,Go87} have 
brought important insight regarding critical phenomena related to 
a collapsing gravitating system. However, deriving the 
differential equations directly from the 4d Einstein equation 
obscures the appearance of the conservation law mentioned above. 
This is probably the reason why -- to the best of our knowledge 
-- it has not been noted and employed in this context before. The underlying
Noether symmetry -- which will not be the issue of this paper -- has been
investigated in \cite{KT98}.

For the reader who is not too familiar with the description of 
spherically reduced Einstein gravity by an equivalent first order 
action with torsion in terms of Cartan variables we recapitulate the action 
and the equations of motion in Section II. 

In Section III we fix the gauge so as to 
produce a diagonal 2d metric. Then the equations used in some recent BH 
simulations \cite{Gu97} are reproduced. For easier comparison our 
notations have been adapted to the ones in that reference. We 
then derive the conservation law in this gauge.

Section IV is devoted to a EF gauge which, in our formulation, 
provides a very similar system of equations. This gauge, which in the 
non-static case is also known as (ingoing) Bondi-Sachs (BS) gauge - c.f. e.g. 
\cite{bondi}, seems to 
be  especially adequate for computations involving both sides of 
the horizon, because it avoids the coordinate singularity, which 
could possibly present problems in the diagonal gauge. Our system of equations
in that gauge resembles closely the one in the diagonal gauge and seems to be 
remarkably simple in comparison with the one used in earlier numerical work 
with BS gauge fixation \cite{MARSA96}. Of course the mathematical content of 
both approaches is identical. Therefore, it must be possible to eliminate the 
extrinsic curvature appearing in \cite{MARSA96}. An explicit example of such 
an elimination in a derivation of the spherically reduced action equation 
(\ref{2}) below has been given in the appendix of the second reference of 
\cite{Thomi84}. Finally a simple relationship between the integrated 
mass-aspect function, the conservation law, the Bondi mass (at ${\cal{I}}^-$) 
and the ADM mass is established.    

In the conclusions we list further applications where the same 
arguments can be applied with equal ease.

\section{First order formulation of spherically reduced gravity}

In terms of the spherically symmetric ansatz for the line element
($g_{\mu \nu}=g_{\mu \nu}(x)=g_{\mu \nu}(t,r)$ with the dilaton field defined 
by $\phi(x)$)
\begin{equation}
\left( ds \right)^2 = g_{\mu \nu}dx^{\mu}dx^{\nu} - 
\frac{4e^{-2\phi}}{\lambda^2} \left( d\Omega \right)^2
\label{1}
\end{equation}
after integrating out the angular variables $ \int d^2\Omega$ of 
$S^2$, the Einstein-Hilbert Lagrangian in $4$ dimensions with 
Newton constant $G=1$ reduces to \cite{Thomi84} a dilaton theory in $d = 
2$:  
\begin{equation}
L^{\left( g \right)}_{dil} = \int d^2x {\cal{L}}^{\left( g \right)}_{dil} 
\text{, } {\cal{L}}^{\left( g \right)}_{dil} = \sqrt{-g} 
e^{-2\phi} \left(R + 2\left( \nabla\phi \right)^2 - 
\frac{\lambda^2}{2}e^{2\phi} \right) \label{2}
\end{equation}

In a similar way the Lagrangian describing minimally coupled 
scalars $S$ in $d = 4$ becomes
\begin{equation}
{\cal{L}}^{\left( m \right)}= 8 \pi \sqrt{-g} e^{-2\phi} \left( 
\nabla S \right)^2 {\text{ .}} \label{3}
\end{equation}

Expressions like $\left( \nabla S \right)^2=g^{\mu \nu} \left( \partial_{\mu} S \right) \left( \partial_{\nu} S \right)$ involve the 2d metric in (\ref{1}). 
Introducing a new variable
\begin{equation}
X = e^{-2\phi}
\label{4}
\end{equation} 
for the dilaton field, the Lagrangian (\ref{3}) can be verified to be 
equivalent to the first order action of a 2d theory with 
nonvanishing torsion\footnote{In (\ref{5}) as well as in all formulae 
derived from it below an overall factor $2$ has been dropped.} \cite{KKL}
\begin{eqnarray}
{\cal{L}}^{\left( g \right)} &=& - \left( X^+D\wedge e^- + X^-D
\wedge e^+ + X d \omega - e^-\wedge e^+ {\cal V}\right) \text{, } \label{5} \\
\text{with } {\cal V} &=& V(X)+X^+X^-U(X) \nonumber \\
\text{where } U(X) &=& -\frac{1}{2X} \text{ and } V(X) = - \frac{\lambda^2}{4}
\text{ for spherically reduced gravity} \nonumber
\end{eqnarray} 
with the covariant derivative $D \wedge e^{\pm} = d \wedge e^{\pm} \pm
\omega \wedge e^{\pm}$.

Equation (\ref{5}) is expressed in terms of Cartan variables, the zweibeine 
$e^a$ and the spin connection $\omega^{ab} = \omega 
\varepsilon^{ab}$. $\varepsilon^{\mu\nu}$ represents the 
antisymmetric Levi-Civit\'a symbol ($\varepsilon^{01} = 
-\varepsilon_{01} = 1$). In the local Lorentz light-cone 
coordinates $ (a = (-,+))$
\begin{equation}
\eta^{ab} = \eta_{ab} = \left( 
\begin{array}{cc}
0 & 1 \\
1 & 0 
\end{array} \right)
\label{6}
\end{equation}
the relation to the 2d metric is
\begin{eqnarray}
g_{\mu \nu} &=& e^a_{\mu} e^b_{\nu} \eta_{ab} \label{7} \\
\sqrt{-g} &=& \left( e \right) = e^-_0e^+_1 - e^-_1e^+_0 {\text{ ,}} \label{8}
\end{eqnarray}
and the 2d curvature scalar becomes
\begin{equation}
\sqrt{-g}R = -2 \varepsilon^{\mu \nu} \partial_{\nu} \widetilde{\omega}_{\mu} {\text{ ,}}
\label{9}
\end{equation}
where $\widetilde{\omega}_{\mu}$ denotes the spin-connection for 
vanishing torsion (i.e. when the term involving $X^\pm$ in (5) 
were absent). The equivalence of (\ref{2}) with (\ref{5}), using (\ref{8}) and 
(\ref{4}) can be checked easily. The algebraic equations of 
motion for $\omega_0$ and $\omega_1$ from the variation with respect 
to $X^\pm$ are reinserted into (\ref{5}), then $X^\pm$ can be 
eliminated in a similar manner.

Now also (\ref{3}) should be expressed by Cartan variables. From the 
identity
\begin{equation}
\sqrt{-g} g^{\alpha \beta} = -\frac{\varepsilon^{\alpha \rho} \varepsilon^{\beta \sigma}}{\left( e \right)} \left( e^+_{\rho} e^-_{\sigma} + e^-_{\rho} e^+_{\sigma} \right)
\label{10}
\end{equation}
in terms of the abbreviations
\begin{equation}
S^{\pm} = \varepsilon^{\alpha \rho} e^{\pm}_{\rho} \partial_{\alpha} S
\label{11}
\end{equation}
equation (\ref {3}) becomes:
\begin{equation}
{\cal{L}}^{\left( m \right)} = - \frac{8 \pi X}{\left( e \right)} S^+ S^-
\label{12}
\end{equation}

The equations of motion (e.o.m.) for the sum of the geometric Lagrangian 
(\ref {5}) and the matter Lagrangian (\ref{12}) are 
\begin{eqnarray} 
dX + X^-e^+ - X^+e^- = 0 \label{13} \\
dX^{\pm} \pm \omega X^{\pm} \mp e^{\pm}\left(V(X)+X^+X^-U(X)
\right) + M^{\pm} = 0 \label{14} \\ 
d \omega - e^- \wedge e^+ \left(V'(X)+X^+X^-U'(X)\right) -
\frac{\delta {\cal{L}}^{(m)}}{\delta X} = 0 \label{18} \\
\left( d \pm \omega \right) \wedge e^{\pm} - e^- \wedge e^+
X^{\pm} U(X) = 0 \label{16} \\
d \left( \frac{F(X)}{(e)}(e^+S^-+e^-S^+) \right) = 0 
\label{19}
\end{eqnarray}
with
\begin{equation}
M^{\pm} = - \frac{\delta {\cal{L}}^{(m)}}{\delta e^{\pm}} 
\end{equation}
using the left-derivative.

In the system of equations (\ref{13})-(\ref{19}) the 2d gauge fixing is still 
free. This is one of the advantages of the approach starting from a spherically
reduced action. The gauge symmetry of (\ref {5}) and (\ref {12}) consists of 
two diffeomorphisms and one local Lorentz transformation ${\cal G} = 
\text{Diff}_2 \times SO(1, 1)$. Thus the equations are not independent. 

In this first order formulation the absolute conservation law is obtained easily \cite{KW95}. Combining 
\begin{eqnarray*}
X^+ \times [eq. (\ref{14})^-] + X^- \times [eq. (\ref{14})^+] \\ 
+ \left(V(X) + X^+X^-U(X)\right) \times [eq. (\ref{13})]
\end{eqnarray*}
where eq. $(\ref{14})^{\pm}$ is the e.o.m. from $\delta e^{\mp}$, we obtain 
with an integrating factor
\begin{equation}
I(X) = exp\left[\int^XU(X')dX'\right]
\end{equation}
(which simplifies to $I(X) = X^{-\frac{1}{2}}$ in the case of spherically 
reduced gravity)
\begin{equation}
d {\cal{C}}^{(g)} + W^{(m)} = 0 \label{cons1}
\end{equation} 
with
\begin{eqnarray}
{\cal{C}}^{(g)} &=& I(X)X^+X^- + \int^X V(X') I(X') dX' \\
W^{(m)} &=& I(X) \left( M^+X^- + M^-X^+ \right) .
\end{eqnarray}

The integrability condition $dW^{(m)}=0$ can be verified to be a consequence of
the e.o.m. (\ref{13})-(\ref{16}) \cite{KT98}, but it also follows trivially 
from the definition (\ref{cons1}). Therefore, with $W^{(m)}=d{\cal{C}}^{(m)}$ 
the conservation law reads $d\left({\cal{C}}^{(g)}+{\cal{C}}^{(m)}\right) = 0$.

For the case of spherically reduced gravity we obtain in components
\begin{equation}
\partial_{\mu} {\cal{C}}^{(g)} + W_{\mu}^{(m)}= 0 \label{21}
\end{equation}
where
\begin{eqnarray}
{\cal{C}}^{\left( g \right)} &=& \frac{X^+X^-}{\sqrt{X}} - \frac{\lambda^2}{2} \sqrt{X} \label{22} \\
W_{\mu}^{\left( m \right)} &=& \frac{8 \pi \sqrt{X}}{\left( e \right)^2} 
\left[ \left( e \right) \left( \partial_{\mu} S \right) \left( S^-X^++S^+X^- 
\right) - S^+S^- \partial_{\mu} X \right] \label{23}
\end{eqnarray}

In the absence of matter ($W^{\left( m \right)} = 0$) 
the quantity ${\cal{C}}^{\left( g \right)} < 0$ is 
proportional to the mass of the black hole; in the presence of matter it 
becomes the mass aspect function (see below). 

\section{Diagonal gauge}

The choice of the gauge for $g_{\mu \nu}$ in (\ref{1}) with equation (\ref{4}) 
(cf. ref. \cite{Gu97}; we use as in (\ref{1}) the opposite 
convention for the sign of $(ds)^2$)
\begin{eqnarray}
g_{\mu\nu} &=& \left( \begin{array}{cc}
\alpha^2 \left( t,r \right) & 0 \\
0 & -a^2 \left( t,r \right) \end{array} \right) \label{37} \\
X &=& \frac{\lambda^2r^2}{4} \label{38}
\end{eqnarray}
for the zweibeine $e_{\mu}^{\pm}$ yields the conditions
\begin{equation}
e^+_0 = e^-_0 = \frac{\alpha}{\sqrt{2}}, \hspace{1cm} e^+_1 = -e^-_1 = 
\frac{a}{\sqrt{2}} . \label{39}
\end{equation}

In the gauge (\ref{37}), (\ref{38}) equation (\ref{19}) is simply solved by 
\begin{equation}
X^+ = X^- = - \frac{\lambda^2r}{2\sqrt{2}a} \text{ .}
\label{40}
\end{equation}

Taking the sum and the difference of equations (\ref{16}) the 
algebraic relations for the spinor connection $\omega_\mu$ are obtained:
\begin{equation}
\omega_0 = \frac{\partial_1 \left( \alpha r \right)}{ar}, \hspace{1cm} 
\omega_1 = \frac{\partial_0 a}{\alpha} \label{41}
\end{equation}

For easier comparison with the equations of motion in ref. \cite{Gu97} 
the same abbreviations 
\begin{equation}
\partial_0 S = \frac{\partial S}{\partial t} = \frac{\alpha}{a} {\cal{\Pi}}, 
\hspace{1cm} \partial_1 S = \frac{\partial S}{\partial r} = {\cal {\Phi}} 
\label{44}
\end{equation}
are introduced. From (\ref{19}) immediately the matter equations 
((5) and (6) of ref. \cite{Gu97}) are reproduced:
\begin{eqnarray}
\partial_0 {\cal{\Pi}} &=& \frac{1}{r^2}
\partial_1 \left( r^2 \frac{\alpha}{a} {\cal {\Phi}} \right) \label{45} \\
\partial_0 {\cal {\Phi}} &=& \partial_1  
\left( \frac{\alpha}{a} {\cal{\Pi}} \right) \label{46}
\end{eqnarray}

The ``Hamiltonian constraint'' (equation (8) of ref. \cite{Gu97}) follows 
from the sum of the equations (\ref{14}) for the component $\mu = 1$: 
\begin{equation}
\partial_1 \ln a + \frac{a^2 - 1}{2r} - 2\pi r \left( {\cal{\Phi}}^2 + 
{\cal{\Pi}}^2 \right) = 0
\label{47}
\end{equation}

Comparing this with the difference of the zero components of the 
same two equations results in the ``slicing condition'' of ref. 
\cite{Gu97}
\begin{equation}
\partial_1 \ln \left( \frac{\alpha}{a} \right) + 
\frac{1 - a^2}{r} = 0 \text{ .} \label{48}
\end{equation}

All other combinations of equations (\ref{14})-(\ref{18}) are found to yield 
dependent equations except the zero component of the sum of equations 
(\ref{14})
\begin{equation}
\partial_0 a = 4 \pi r \alpha {\cal{\Phi}}{\cal{\Pi}} \text{ ,}
\label{49}
\end{equation}
which, as will be seen below, in a certain sense contains further information.

Having verified that the basic equations (\ref{45})-(\ref{48}) are identical 
to the ones following from the 4d Einstein equations\footnote{Although 
seemingly obvious at first glance this result is by no means trivial. E.g. for 
the more general ansatz of a ``warped'' metric in Einstein relativity 
\cite{KKK98} the complete set of e.o.m. cannot be derived from the reduced 
action. Also the treatment of quantum effects like Hawking radiation cannot be 
carried over straightforwardly to the spherically reduced case \cite{MWZ94}.} 
in we turn to the absolute conservation law (\ref{21}). In the gauge 
(\ref{37}), (\ref{38}) its geometric part becomes
\begin{equation}
{\cal{C}}^{\left( g \right)} = \frac{\lambda^3r}{4a^2} \left( 1 - a^2 \right)
\label{50}
\end{equation}
which in terms of a variable mass parameter $m(t,r)$, the so-called mass
aspect function \cite{Gu97}
\begin{equation}
a^{-2} = 1 - \frac{2m}{r}
\label{51}
\end{equation}
yields
\begin{equation}
{\cal{C}}^{\left( g \right)} = - \frac{\lambda^3}{2} m \text{ ,}
\label{52}
\end{equation}
i.e. in the absence of matter ${\cal{C}}^{\left( g \right)}$ is proportional 
to the mass of the (Schwarz\-schild) black hole.

The matter contribution to (\ref{21}) in the present gauge with 
(\ref{44}) becomes
\begin{eqnarray}
\partial_0 {\cal{C}}^{\left( m \right)} &=& -\frac{\lambda^3}{2} \rho^2 \left( 
2\frac{\alpha}{a}{\cal {\Pi\Phi}} \right) 
\label{53} \text{ ,} \\
\partial_1 {\cal{C}}^{\left( m \right)} &=& -\frac{\lambda^3}{2} \rho^2 
\left( {\cal{\Phi}}^2 + {\cal{\Pi}}^2  \right) 
\text{ .} \label{54}
\end{eqnarray}

With (\ref{52}) and the abbreviations \cite{Gu97}
\begin{equation}
{\cal{X}} = \rho{\cal \Phi}, \hspace{1cm} {\cal{Y}} = \rho{\cal \Pi}, 
\hspace{1cm} \rho = \frac{\sqrt{2\pi}r}{a} \label{55}
\end{equation}
the conservation law (\ref{21}) can be brought into a very simple form:
\begin{eqnarray}
\partial_0 m &=& 2 \frac{\alpha}{a}{\cal{X}}{\cal{Y}} = A \left( t,r \right) 
\label{57} \\
\partial_1 m &=& {\cal{X}}^2 + {\cal{Y}}^2 = B \left( t,r \right) \label{58}
\end{eqnarray}

Clearly in the present gauge these equations are nothing else but 
the equations of motion (\ref{47}) and (\ref{49}). Therefore in this 
particular case the whole formalism leading to (\ref{21}) seems to be 
superfluous. However, as emphasized already above, the power of 
that approach becomes evident in other gauges, e.g.\ when the dilaton field
$X=X(t,r)$ is not included among the gauge fixed quantities. 

The integral of (\ref{57}) and (\ref{58}) can be written in two ways. 
Starting with the time integral of (\ref{57}) and then using (\ref{58}) 
with the integrability condition for the r.h.s. of these two 
equations yields
\begin{equation}
m \left( t,r;t_0,r_0 \right) = \int^t_{t_0} dt' A \left( t',r \right) + \int^r_{r_0} dr' B \left( t_0,r' \right) + m_0
\label{59}
\end{equation}
which determines $m$ for arbitrary $t,r$ from boundary values 
$r_0$ and $t_0$ and from the dynamical evolution of the scalar 
fields. The overall conserved value is represented by the 
constant $m_0$ which labels a certain solution much like the 
total energy in an ordinary conservative mechanical system. An 
equivalent formulation of the integral (\ref{59}) is
\begin{equation}
m \left( t,r;t_0,r_0 \right) = \int^r_{r_0} dr' B \left( t,r' \right) + \int^t_{t_0} dt' A \left( t',r_0 \right) + m_0 \text{ .}
\label{60}
\end{equation}

Equation (\ref{59}) can be transformed directly into (\ref{60}) by the use of 
the integrability condition for the r.h.s. of (\ref{57}) and (\ref{58}). 
It can be seen from (\ref{59}) or (\ref{60}) that equation (\ref{57}), which 
is the same as (\ref{47}), contains new information. Only the integrability 
condition involving its r.h.s.\ depends on the minimal set of 
equations. The new information is  the one leading to the 
conserved quantity $m_0$. 

It is instructive to extract the usual ADM mass \cite{ADM} from (\ref{59}). In
the limit $r \to \infty$ for a system with an asymptotic Killing field 
$\frac{\partial}{\partial t}$ the ADM mass follows from (\ref{59})
\begin{equation}
m_{ADM} = \int^{\infty}_0 dr' B \left( t_0 ,r' \right) + m_0 \label{61}
\end{equation}
since $\frac{\partial m}{\partial t}=0$ requires vanishing $A(t,\infty)$. In
(\ref{61}) the lower limit of the integral has been changed with a simultaneous
redefinition of $m_0$. 

With the asymptotic spherical wave solutions in $A$ ($u = t-r, v=t+r, a \to 1,
\alpha \to 1$)
\begin{equation}
S = \frac{1}{\sqrt{4\pi}r}\left(f_+(u)+f_-(v)\right) \label{62}
\end{equation}
from (\ref{44}), (\ref{55}), (\ref{57}) and (\ref{58}) the mass aspect 
function in the limit $r \to \infty$ becomes
\begin{equation}
m_{BH}^{eff} (t) = m (t, \infty; t_0, 0) = m_{ADM} - \int_{t_0}^{t} dt' \left[
\left(f'_+\right)^2-\left(f'_-\right)^2\right], \label{63}
\end{equation}
to be interpretated as the total effective mass of the (eventual) black hole.
It consists of the usual ADM mass minus the difference of total outgoing and
ingoing fluxes of matter at a certain time $t$. It is remarkable that a time
dependent Bondi-like mass appears without having used the (for this purpose
traditional) BS-gauge (see below). If in- or outgoing matter fluxes exist it is
neccessary to use the effective black hole mass $m_{BH}^{eff}$ as a measure 
for black hole formation rather than $m_{ADM}$ alone \cite{Chop86}.

\section{Bondi-Sachs gauge}

The advantage of the (ingoing) Eddington-Finkelstein gauge for the 2d metric 
$g_{\mu\nu}$ in (\ref{1}), defined either by $(x^0 = v, x^1 =r)$
\begin{equation}
\left( ds \right)^2 = (e) \left[ h dv - 2 dr \right] dv \label{64}
\end{equation}
or by
\begin{equation}
e^+_1 = 0 \text{ , } \hspace{1cm} e^-_1 = -1
\label{65}
\end{equation}
and
\begin{equation}
e^+_0 = \det e = \left( e \right) = \sqrt{-g} \text{ , } 
\hspace{1cm} e^-_0 = \frac{h}{2} 
\label{66}
\end{equation}
is its regular behavior at the event horizon. 
The Killing norm $(e)h$ for the Schwarzschild black hole has a 
simple zero. For this reason this 
gauge has been used also previously within studies of black hole 
formation (e.g.\ \cite{MARSA96}, \cite{Chri84}).  In the presence of matter
$(e)$ and $h$ are functions of $r$ and $v$ (ingoing BS gauge). In the first 
order approach in equations (\ref{13})-(\ref{19}) now the gauge is fixed by 
(\ref{65}) and (\ref{38}). 

Again (\ref{14}) determines
\begin{equation}
X^+ = -\frac{\lambda^2}{2}r, \hspace{1cm} X^- = -\frac{h}{4 (e)}\lambda^2r 
\label{67}
\end{equation}
and as in the diagonal gauge the equations (\ref{16}) define the spin 
connection: 
\begin{equation}
\omega_0 = \frac{\partial_1 h}{2} + \frac{h}{2} \frac{\partial_1 (e)}{(e)} + 
\frac{h}{r}, \hspace{0.5cm} \omega_1 = -\frac{1}{r} - \frac{\partial_1 (e)}
{(e)} \label{69}
\end{equation}

From (\ref{14}) for $\mu = 1$ with (\ref{69}) the BS-analogue of the 
``slicing condition'' (\ref{48}) of the diagonal gauge is obtained:
\begin{equation}
\partial_1 \ln h + \frac{1}{r} \left(1 - \frac{(e)}{h} \right) = 0
\label{70}
\end{equation}

It also does not involve derivatives $\partial_0$
and is even simpler than in the diagonal gauge. The ``Hamiltonian constraint'' 
\begin{equation}
\frac{r}{2} \partial_1 \ln \left( e \right) - \hat{{\cal{X}}}^2 = 0
\label{71}
\end{equation}
is a consequence of (\ref{14}) with $\mu = 1$ and (\ref{69}). In a similar way 
the analogue of equation (\ref{49})
\begin{equation}
\partial_0 \left( r \frac{h}{\left( e \right)} \right) - r \frac{h^2}{2} 
\frac{\partial_1 (e)}{(e)^2} + \frac{4}{\left( e \right)} 
\hat{{\cal{Y}}}^2 = 0 \label{72}
\end{equation}
follows with the definitions
\begin{eqnarray}
\hat{{\cal{X}}} &=& \sqrt{2\pi}r\partial_1 S \label{73} \\
\hat{{\cal{Y}}} &=& \sqrt{2\pi}r \left( \frac{h}{2} \partial_1 S + \partial_0 S \right) \label{74}
\end{eqnarray}

The e.o.m. (\ref{19}) for the scalar field in terms of (\ref{73}) and 
(\ref{74}) become
\begin{equation}
r \partial_1 \hat{{\cal{Y}}} + \frac{h}{2} \hat{{\cal{X}}} = 0 \label{75}
\end{equation} 
and 
\begin{equation}
\left( \partial_0 - \partial_1 \frac{h}{2} \right) \frac{\hat{{\cal{X}}}}{r} = - \partial_1 \left( \frac{\hat{{\cal{Y}}}}{r} \right) \text{ .}
\label{76}
\end{equation}

The geometric part of the conserved quantity (\ref{22}) with (\ref{67}) yields
\begin{equation}
{\cal{C}}^{\left( g \right)} = \frac{\lambda^3r}{4} \left( \frac{h}{\left( e \right)} - 1 \right)
\label{77}
\end{equation}
which by comparison with the result (\ref{52}) allows the introduction 
of a mass aspect function $\hat{m}(v, r)$. Evaluating $W_0$ and $W_1$ in 
(\ref{23}) the absolute conservation law for $\hat{m}$ reads
\begin{eqnarray}
\partial_0 \hat{m} &=& \frac{2}{(e)} \left({\hat{{\cal{Y}}}}^2 - \frac{h^2}{4}
{\hat{{\cal{X}}}}^2 \right) = \hat{A}(v, r) \text{ ,} \label{78} \\
\partial_1 \hat{m} &=& \frac{h}{(e)} {\hat{{\cal{X}}}}^2 = \hat{B}(v, r) 
\text{ ,} \label{79}
\end{eqnarray}
upon which a similar argument for an integrated mass-function as the one from 
(\ref{57}) and (\ref{58}) can be based. 

Performing analogous steps as in the diagonal gauge we obtain in the limit of 
$\lim_{r \to \infty}$ the Bondi-mass at ${\cal{I}}^-$
\begin{equation}
\hat{m}_-(v) = \lim_{r \to \infty} \hat{m}(v, r) = \hat{m} (v, \infty; \infty,
0) = \int_{\infty}^{v} dv' \hat{A}(v', \infty) + \hat{m}_{ADM} \label{bondi}
\end{equation}
and similarly the ADM-mass
\begin{equation}
\hat{m}_{ADM} = \hat{m}_-(\infty) = \hat{m} (\infty, \infty; \infty, 0) = 
\int_0^{\infty} dr' \hat{B}(\infty, r') + \hat{m}_0 . \label{ADM}
\end{equation}

Of course, a similar set of equations is obtained for asymptotically 
($h \to 1$, $(e) \to 1$) outgoing BS gauge
\begin{equation}
(ds)^2 = (e) \left( hdu+2dr \right)du \label{100}
\end{equation}
instead of (\ref{64}), with $h$ and $(e)$ now being functions of $u$ and $r$
leading to the Bondi mass $m_+(u)$ at ${\cal{I}}^+$.

The simple set of equations (\ref{70}) and (\ref{71}) should be compared with 
the equations to be used when the extrinsic curvature has not been eliminated 
\cite{MARSA96}. This explains the difficulty to pinpoint in that work the 
conservation law (\ref{78}) and (\ref{79}) and its ensuing integrated form as 
in (\ref{59}) and (\ref{60}).

\section{Conclusions and Outlook}

We emphasize the usefulness of a first order formulation of 
spherically reduced gravity. It allows in a very easy manner to 
specify particular gauges from a general set of first order 
differential equations, involving Cartan variables and auxiliary 
fields, one of which can be identified with the dilaton field.

However, the central point of our argument is that the (in space and time) absolute conservation law, valid for all 2d theories of gravity should be 
properly taken into account in treatments of selfgravitating 
matter because it directly produces a time dependent effective black hole mass
(\ref{63}). 

It is instructive to compare our result with the one obtained in 
the seminal work of Mann \cite{mann}. There are similarities in the 
structure of the resulting relations, but also essential 
differences which allow us to cover a much wider field of 
applications than Mann's formula is able to do. Whereas kinetic 
terms from scalars (and also for fermions), to be used in 
computer simulations as e.g.\ in \cite{Chop86}, \cite{MARSA96}
fit perfectly in our 
generalized version of that conservation law, those physically 
important cases had to be excluded explicitly in \cite{mann}\footnote{Cf.\ the 
remark after eq. (4) of ref \cite{mann}.}. Apart from 
that also our conservation law is obtained in a first stage by 
combining the e.o.m. (\ref{13}), (\ref{14}) for {\em auxiliary} variables $X, 
X^\pm$ in order to arrive at the relation (\ref{cons1}) which only without 
matter ($W^{(m)} = 0$) would correspond to the conservation of ${\cal 
C}^{(g)}$ which essentially coincides with the ADM-mass. Then, as 
in \cite{mann} by using in addition the ``genuine'' e.o.m. from the 
Einstein-Hilbert action with matter (eqs.\ (\ref{18})-(\ref{19}) in our paper), 
one can prove the integrability condition\footnote{Cf.\ the 
second reference \cite{KT98}. Of course, the validity of $dW^{(m)} = 0$ 
also follows trivially from the structure of (\ref{cons1}).} 
$dW^{(m)} = 0$ and thus a conservation for the {\em sum} of 
${{\cal C}^{(g)}}$ and ${{\cal C}^{(m)}}$. \\
In fact, this ``two-stage'' structure of the conservation law is 
reflected in the associated Noether-symmetry \cite{KT98}.  Again a 
formulation in terms of Cartan variables has turned out to be 
most illuminating. Whereas in the matterless case the 
Noether-symmetry reduces to the well-known transformations in the 
direction of the Killing-field (called $\delta\gamma$ in the 
second reference \cite{KT98}), in the presence of matter the 
integrability condition $d W^{(m)} = 0$, referred to above, must 
be interpreted as ``another'' conservation law for a 1-form 
``current'' $W^{(m)}$ with associated matter-related symmetry 
parameters ($\delta \rho$ in the second reference \cite{KT98}) which are 
{\em different} from the $\delta\gamma$. 

In the matterless case there is even a deeper reason for the existence of the 
conserved quantity: Since the geometric part of the Lagrangian (\ref{5}) is a 
special case of a Poisson-$\sigma$ model \cite{Str94},\cite{Scha94}, one can 
use so-called Casimir-Darboux coordinates in order to derive the (geometric 
part of the) conservation law (\ref{cons1}). For the reader not so familiar 
with Poisson-$\sigma$ models and its ensuing relation to first order gravity 
we refer to \cite{KS96}.

In our present paper we discuss the conservation law including matter for
the diagonal gauge \cite{Chop86,Gu97} and for the Bondi-Sachs gauge 
\cite{bondi}.

Comparing for the diagonal gauge the  
equations used for numerical simulations for black hole formation 
we find that the conservation law is intimately related to a well-known 
differential equation of the time derivative of the 
mass aspect function $m(t,r)$ in the diagonal 
gauge. The conservation law for $m = m(t,r)$ is formulated in terms of an 
integral over the scalar fields and an additional free 
constant $m_0$ which may be interpreted as the initial value of 
the geometric part of the action. The latter, together with the 
initial value of the scalar field through (\ref{61}) determines the solution 
and is identical with the ADM mass. The mass aspect function at $i_0$, eq.
(\ref{63}), represents the effective black hole mass at a certain time $t$.
It depends on the in- and outgoing matter fluxes thus leading to
a Bondi-like mass definition already in a diagonal gauge. 

In the Bondi-Sachs gauge we obtain a remarkably simple relationship between
the Bondi mass, the ADM mass, the conservation law and the mass aspect 
function, summarized in eqs. (\ref{bondi}) and (\ref{ADM}). Also, the other
e.o.m.'s are much simpler than the ones in some recent literature
\cite{MARSA96}, since we are able to avoid altogether the introduction of the 
extrinsic curvature as a dynamical variable.

Numerous further applications of our present approach are 
obvious: 

Spherically reduced Einstein-gravity in $d > 4$ dimensions in the 
line element (1) only  shows a different power of the dilaton field\footnote{
The factor of $\left(d\Omega\right)^2$ has been chosen such that the dilaton 
field is dimensionless and in the limit $\lim_{d \to \infty}$ we obtain the 
CGHS model (c.f. fourth reference of \cite{MAN91}) for some finite $\lambda$
of mass dimension one.}
\begin{equation}
\left( ds \right)^2 = g_{\mu \nu}dx^{\mu}dx^{\nu} - \frac{(d-2)^{\frac{4}
{(d-2)}}}{\lambda^2} e^{-\phi \frac{4}{(d-2)}} \left( d\Omega \right)^2
\label{80}
\end{equation}
which leads to a replacement of the ``potential'' ${\cal V}$ in the geometric 
part of the action (\ref{5}) by
\begin{equation}
{\cal V}^{\left( d \right)} = - \frac{(d - 3)}{(d - 2)} \frac{X^+X^-}{X} -
\frac{\lambda^2}{2} \frac{(d - 3)}{(d - 2)} X^{\frac{d-4}
{d-2}} \text{ .} \label{82}
\end{equation}

The integrating factor in the conservation law changes from $X^{-\frac{1}{2}}$ 
to $X^{-\frac{d-3}{d-2}}$. With the gauge fixing 
$X = \frac{(\lambda r)^{(d-2)}}{4}$ all further steps are exactly as in the 
case $d = 4$.

The conservation law (\ref{21}), of course, also appears when other 
types of matter are considered. In conformal gauge (cf.\ e.g.\ 
\cite{Fro97}) for a 2d line element $(ds)^2 = 2 e^{2\rho} du\, dv$  
the proper gauge in (1) is $e^-_u = e_v^+ = 0, e_u^+ = e_v^-  = 
e^\rho$.  

Other gauges may be chosen with equal ease. For 
instance the area $A(t,r) = 8\pi X/\lambda^2$ of the surface 
$S^2$ could be retained as an independent variable, and, say, in 
$g_{\mu\nu}$ the BS gauge could be restricted further by 
requiring the 2d volume $\det (e) = 1$ as in the Schwarzschild gauge. 

So far only spherically reduced Einstein gravity has been considered. The 
generalization to the Einstein deSitter case is obvious as well. A 
nonvanishing cosmological constant $\Lambda$ simply changes the ``potential'' 
${\cal V}$ to
\begin{equation}
{\cal V}^{\left( d=4 \right)}_{deSitter} =  - \frac{X^+X^-}{2X} - \frac{\lambda^2}{4} + \Lambda X \text{ .} 
\label{83}
\end{equation}
Then, e.g. in the diagonal gauge of section 3, neither the e.o.m. for 
(\ref{45}), (\ref{46}) for the scalar field, nor the conservation law 
equations (\ref{57}), (\ref{58}) are changed, if the mass aspect function is 
redefined as
\begin{equation}
a^{-2} = 1 - \frac{2m \left( t, r \right) }{r} - \frac{\Lambda r^2}{3} \text{ .}
\label{84}
\end{equation}

Only the ``slicing-condition'' (\ref{47}) acquires an additional term
\begin{equation}
\frac{\partial}{\partial r}\ln \left( \frac{\alpha}{a} \right) + \frac{1 - a^2}{r} + \frac{\Lambda a^2 r}{2} = 0 \text{ ,}
\label{85}
\end{equation}
as well as the ``Hamiltonian constraint'' (\ref{47})
\begin{equation}
\frac{\partial}{\partial r}\ln a + \frac{a^2 - 1}{2r} - \frac{\Lambda a^2 r}{2} = 2\pi r \left( {\cal{\Phi}}^2 + {\cal{\Pi}}^2 \right) \text{ .}
\label{86}
\end{equation}

Different types of matter can be discussed 
either by inserting the appropriate (spherically reduced) matter 
action or, more directly, from the corresponding (spherically 
reduced) energy momentum tensor $T^{\mu\nu}$. Then instead of the 
$S$-dependent terms the matter interaction in (\ref{14}) is 
proportional to $(e) \varepsilon_{\rho\mu} T^{\rho\sigma} 
e^\pm_\sigma$.  This would be needed e.g.\ for 
the case of a perfect fluid which also has been studied in 
connection with black hole formation \cite{Koike95}. One of the 
main advantages of the 2d spherically reduced gravity formulation 
is the ease by which a transition from one 2d gauge to another is 
possible and the ease by which the absolute conservation law (\ref{21}) 
can be derived. We emphasize once again that the appearance of the latter is 
peculiar to $d = 2$ (or to a 2d reduction of a higher dimensional theory).

Finally it should be mentioned that the conservation law in the 
presence of matter is related to a Noether-symmetry of a
somewhat unusual type \cite{KT98}. This (nonlinear) 
symmetry so far is known only in infinitesimal form. Possibly 
further investigations in that direction could also contribute 
towards the understanding of the intriguing results from 
numerical studies of spherical black hole formation.

\section*{Acknowledgements}

The authors have benefitted from discussions with H. Balasin, D. Hofmann, 
W. Waltenberger and R. Wimmer. They thank M. W. Choptuik for correspondence. 
This research has been supported by Project P12815-TPH of the Austrian Science
Foundation (\"Osterreichischer Fonds zur F\"orderung der wissenschaftlichen 
Forschung).

\vfil

\end{document}